# Quantum Imaging via Kurtosis-Difference Weighted Covariance on a 2D Camera

Zhe He,*,# Yanli Shi,# Hui Wu, Qun Cao, Weidong Zheng, and Zheng Cui

**ABSTRACT:** Camera-based quantum imaging detects spatially correlated photon pairs from spontaneous parametric down-conversion (SPDC). Conventional covariance methods typically require tens of thousands of frames to extract weak correlations from noise. While thick crystals can increase photon flux, they generate photon pairs from multiple emission positions within the crystal, producing multiple correlation centers with complex pairing geometries. In addition, conventional covariance methods assume a single preselected correlation center and cannot fully exploit these distributed correlations. We demonstrate that kurtosis difference, a fourth-order statistic measuring tail similarity, effectively discriminates correlated pixel pairs even when correlation coefficients remain low.

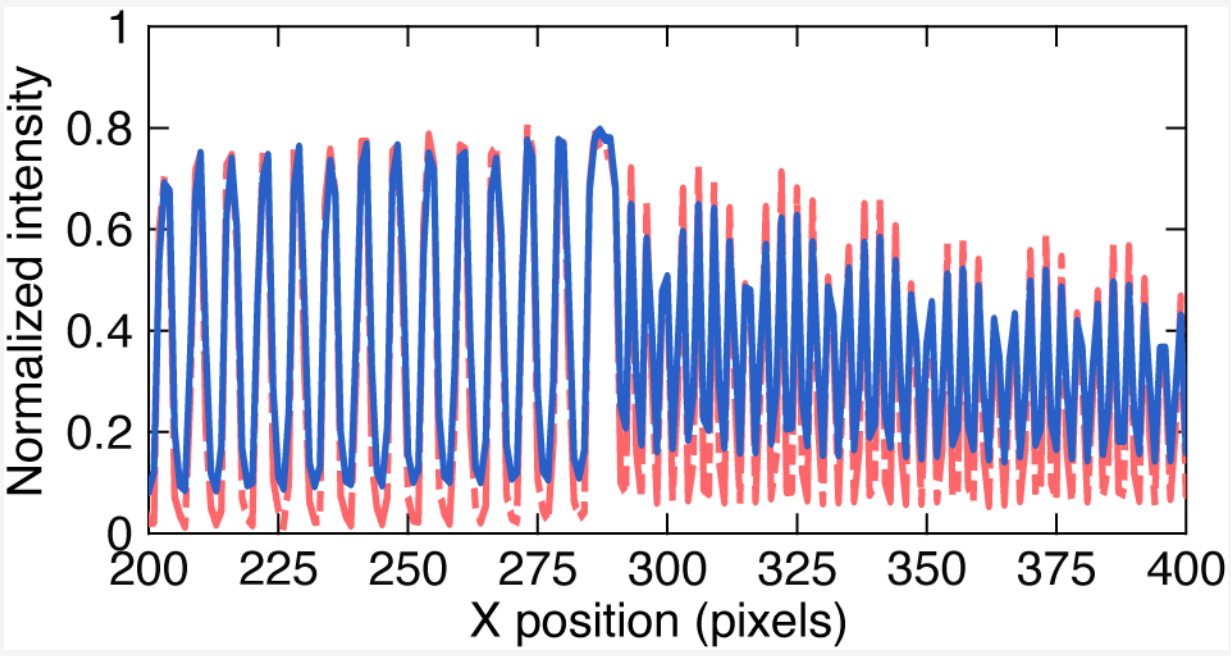

Weighting covariance by an exponential function of absolute kurtosis difference can select symmetric pixels while preserving true coincidences. This kurtosis weighting automatically identifies correlated pairs within a broad search region and accommodates multiple pairing geometries without requiring precise correlation center calibration. At 5000 frames, our method yields a contrast-to-noise ratio (CNR) exceeding 7, whereas standard covariance remains below 2. Compared with standard covariance, the method reduces the acquisition time by 40-fold and could enable practical quantum imaging in sparse correlated-photon regimes.

## ■ INTRODUCTION

Quantum imaging with spatially entangled photon pairs from spontaneous parametric down-conversion (SPDC) outperforms classical imaging in spatial resolution and signal-to-noise ratio.[1−5] Camera based detection captures spatial correlations across many pixel pairs simultaneously, enabling parallel detection without temporal coincidence circuits.[6,7] However, cameras integrate photons over an exposure window. Within this time window, noise, accidental coincidences, and detector imperfections can degrade the data.[8] As a result, extracting weak spatial correlations from camera measurements remains difficult.

Most camera-based approaches estimate correlations of pixel intensities, which reflects second-order statistics. However, the methods usually perform poorly at low photon flux and require tens of thousands of frames to accumulate sufficient statistics.[3] This requirement restricts quantum imaging to stable samples, long acquisition times, and adequate illumination.[9] Thick crystals can increase the photon flux, but they also produce photon pairs from multiple emission positions inside the crystal. The resulting correlations are distributed over multiple centers instead of the single symmetric pairing geometry assumed in conventional covariance analysis.[10] Multicenter strategies can partly address this problem, but they require preselecting correlation centers from prior knowledge of the pairing geometry and may therefore miss valid correlated pairs.[11] To date, a method that can identify correlated pairs automatically, without geometric preselection, is still needed.

SPDC pair counting is usually a rare event.[12] Upon creation of a pair, both photons reach the detector in a single exposure interval and cause correlated intensity spikes. These rare events appear in the tails of pixel intensity distributions.[13] Correlated pixels detecting photons from the same SPDC emission position have similar rare-event statistics and thus similar distribution tails. Uncorrelated pixels, which detect photons from different SPDC events within the crystal, have different distribution tails because their photon sources are independent. To overcome these, we use absolute kurtosis difference as a measure of tail compatibility between pixel pairs. A small difference indicates similar rare-event statistics characteristic of correlated pixels, while a large difference signals independent noise. Absolute kurtosis difference as a weighting criterion discriminates correlated from uncorrelated pixel pairs based on

Received: August 1, 2026
Revised: October 14, 2026
Accepted: October 14, 2026

tail statistics. This provides a practical tool for extracting spatial correlations in SPDC imaging. The method does not require quantum specific signatures or serve as an entanglement witness. It identifies pixel pairs sharing similar sparse photon statistics as the source produces rare correlated events with similar heavy-tailed intensity distributions in symmetric pixels.

We weigh each pixel pair's covariance by absolute kurtosis difference. This exponential weighting evaluates all candidate pixel pairs within a broad search region based on their tail statistics rather than geometric constraints, eliminating the need for preselected correlation centers. Because the criterion is statistical rather than geometric, it can accommodate multiple pairing geometries and correlation structures caused by thick crystals or optical aberrations. We validate this experimentally using Type-I SPDC with a $\beta$-BBO crystal. With 5000 frames, the kurtosis-weighted reconstruction achieves performance comparable to or better than standard covariance reconstruction with 200000 frames. This 40-fold reduction in frame number may enable more practical camera-based quantum imaging in low-flux regimes.

## THEORY

Let $I_i(t)$ denote the measured intensity of pixel $i$ in frame $t$,

$$I_i(t) = G_i(t) + D_i(t) + \eta_i \sum_m P_{m,i}(t) \tag{1}$$

where $G_i(t) \sim \text{Norm}(\mu_{G,i}, \sigma^2_{G,i})$ represents Gaussian noise from electronic detector and optical background, $D_i(t) \sim \text{Pois}(\lambda_{D,i})$ denotes uncorrelated Poisson noise from dark counts and stray photons, $\eta_i$ is the local quantum efficiency, and $P_{m,i}(t) \sim \text{Pois}(\lambda_{m,i})$ represents photon counts from SPDC emission position m within the crystal detected at pixel $i$.

Different emission positions $m$ within the nonlinear crystal generate independent SPDC pairs. Each position $m$ produces entangled photon pairs detected at symmetric pixels $i$ and $j$, where $j = C_m(i)$ is determined by phase-matching geometry. For photons from the same emission position $m$, both photons arrive simultaneously with the same expected intensities

$$\begin{aligned} P_{m,i}(t) &= P_{m,j}(t) \\ \lambda_{m,i} &= \lambda_{m,j} \end{aligned} \tag{2}$$

when $j = C_m(i)$. Kurtosis measures the heaviness of distribution tails through the standardized fourth moment. For pixel $i$, the (excess) kurtosis $\kappa_i$ is

$$\kappa_i = \frac{\langle (I_i - \langle I_i \rangle)^4 \rangle}{\sigma_i^4} - 3 \tag{3}$$

where $I_i$ is the temporal mean, $\sigma_i = \sqrt{\langle I_i^2 \rangle - \langle I_i \rangle^2}$ is the standard deviation of the pixel intensity, and $\langle \cdots \rangle$ denotes temporal average over $T$ frames. The kurtosis of a Poisson distribution with rate $\lambda$ is $\kappa = 1/\lambda$. Thus, the expected kurtosis scales approximately as $\langle \kappa_i \rangle \approx 1/\lambda_i$ when $\lambda_i$ is the dominant photon source for pixel $i$. The kurtosis difference

$$\Delta\kappa_{ij} = \kappa_i - \kappa_j \tag{4}$$

has zero mean for both symmetric and asymmetric pixels: $\langle \Delta\kappa_{ij} \rangle_{\text{pixel}} = 0$, where $\langle \cdots \rangle_{\text{pixel}}$ denotes expected average over multiple pixels (see Supplementary Note 1). However, the variance behaves distinctly.

For symmetric pairs, both pixels receive photons from the same emission position $m$. When $m$ represents the source of an entangled photon pair, both pixels register simultaneous counts. This synchronous fluctuation creates strong covariance between $\kappa_i$ and $\kappa_j$. The variance decomposes as

$$\text{Var}(\Delta\kappa_{ij}) = \text{Var}(\kappa_i) + \text{Var}(\kappa_j) - 2\text{Cov}(\kappa_i, \kappa_j) \tag{5}$$

For symmetric pairs, the positive covariance term nearly cancels the individual variances:

$$\text{Var}(\Delta\kappa_{ij})_{\text{sym}} \ll \frac{1}{T} \tag{6}$$

For asymmetric pixels, pixels $i$ and $j$ sample photons from different emission positions. Their kurtosis values fluctuate independently with $\text{Cov}(\kappa_i, \kappa_j) = 0$:

$$\text{Var}(\Delta\kappa_{ij})_{\text{asym}} = \text{Var}(\kappa_i) + \text{Var}(\kappa_j) \approx \frac{48}{T} \tag{7}$$

In Supplementary Note 1, we derive eq 6 and that the variance of kurtosis for a Poisson distribution is $24/T$. A simulation of this variance is shown in Figure S1.

We define the absolute kurtosis difference as $\delta = |\Delta\kappa_{ij}|$. The expected value of $\delta$ is proportional to $\sqrt{\text{Var}(\Delta\kappa_{ij})}$ (see Supplementary Note 2):

$$\langle \delta \rangle_{\text{pixel}} \approx \sqrt{\frac{2}{\pi}\text{Var}(\Delta\kappa_{ij})} \tag{8}$$

Therefore, small $\delta$ indicates symmetric pairs, while large $\delta$ indicates independent noise. The kurtosis difference measures similarity of fourth-order marginal structure, fundamentally distinct from covariance, which measures joint synchronous fluctuations. The simulation results are shown in Figure S1.

In low-flux SPDC imaging, rare frames with photon pairs produce high intensities that dominate the fourth moment. Correlated pixels detecting photons from the same SPDC events develop similar heavy-tailed distributions and similar kurtosis. We combine this with covariance reconstruction. For each image pixel $i$, we weight the covariance that captures correlated photon counts based on kurtosis difference. The reconstructed intensity is

$$I_{\text{recon},i} = \sum_j \text{Cov}(I_i, I_j) w_{ij} \tag{9}$$

where $j$ scans across every remaining pixel $j \neq i$, and the weight is

$$w_{ij} = \exp\left(-\frac{\delta}{\kappa_0}\right) \tag{10}$$

Here, the $\text{Cov}(I_i, I_j)$ can identify the true coincidence between pixels $(i, j)$. The weight $w_{ij}$ provides data-driven pair selection. We use the absolute kurtosis difference $\delta$ to provide symmetric treatment irrespective of whether $\kappa_i$ or $\kappa_j$ is larger. Small $\delta$ preserves the covariance contribution, whereas large $\delta$ exponentially suppresses it. The parameter $\kappa_0$ defines a characteristic scale for kurtosis difference. It specifies a threshold where the exponential factor reduces the weight by a factor of $e^{-1} \approx 0.37$. For $\kappa_0 \to \infty$, standard covariance is recovered. With finite $\kappa_0$, pixels with both covariance and tail consistency are preferred. Here we choose $\kappa_0 = 1$ so that correlated pairs ($\delta \sim \kappa_0$) are only moderately downweighted, while uncorrelated ones ($\delta \gg \kappa_0$) are strongly attenuated.

Covariance methods assume a single, preselected correlation center and enforce central-symmetry pairing. This assumption breaks down when optical aberrations or thick crystals create multiple pairing centers or distorted symmetry.[10,14,15] Figure S2 demonstrates this challenge: the accumulated intensity (Figure S2a) shows the SPDC illumination pattern, while the correlation map (Figure S2b) reveals multiple distinct pairing centers arising from different emission positions within the crystal. When a single-center assumption is imposed on such multicenter data, photons from different emission positions are incorrectly paired, mixing true correlations with accidental coincidences and degrading reconstruction quality.[16] Increasing number of potential correlation centers helps achieve a better reconstruction quality, but at the cost of substantially increased difficulty in finding the correct center.[11] Rather than exploring a limited number of correlation centers, our method evaluates all candidate pairs and applies kurtosis weighting to distinguish symmetric from asymmetric pixels. This eliminates the need for precise calibration and improves discrimination between true SPDC correlations and noise.

## METHOD

Type-I SPDC was achieved by focusing a CW laser (405 nm, 100 mW) into a $\beta$-BBO crystal (Figure 1). Degenerate down-converted photons at 810 nm went through an optical chain equipped with a USAF resolution target placed along the signal path. Imaging of transmitted light was done with an Andor iXon Ultra 888 EMCCD camera with 13 $\mu$m pixel pitch operated at −90 °C and exposure time of 10 ms with 2 × 2 binning. The EMCCD's high single-photon sensitivity (quantum efficiency above 90% at 810 nm) and low read noise make it ideal for camera-based quantum imaging. In the illumination region, pixels received less than one photon per frame on average, thus the experiment was conducted in a low-flux photon regime characterized by rare pair-generation events dominating the intensity statistics. Data sets consisted of 5000−200000 frames. Periodic bar resolution targets containing bar groups at low, medium, and high frequency are also used for resolution characterization. All periodic-target reconstructions use 45000 frames. The baseline was selected from a corner of the image without SPDC illumination, which was subtracted from all acquired raw images prior to statistical analysis.

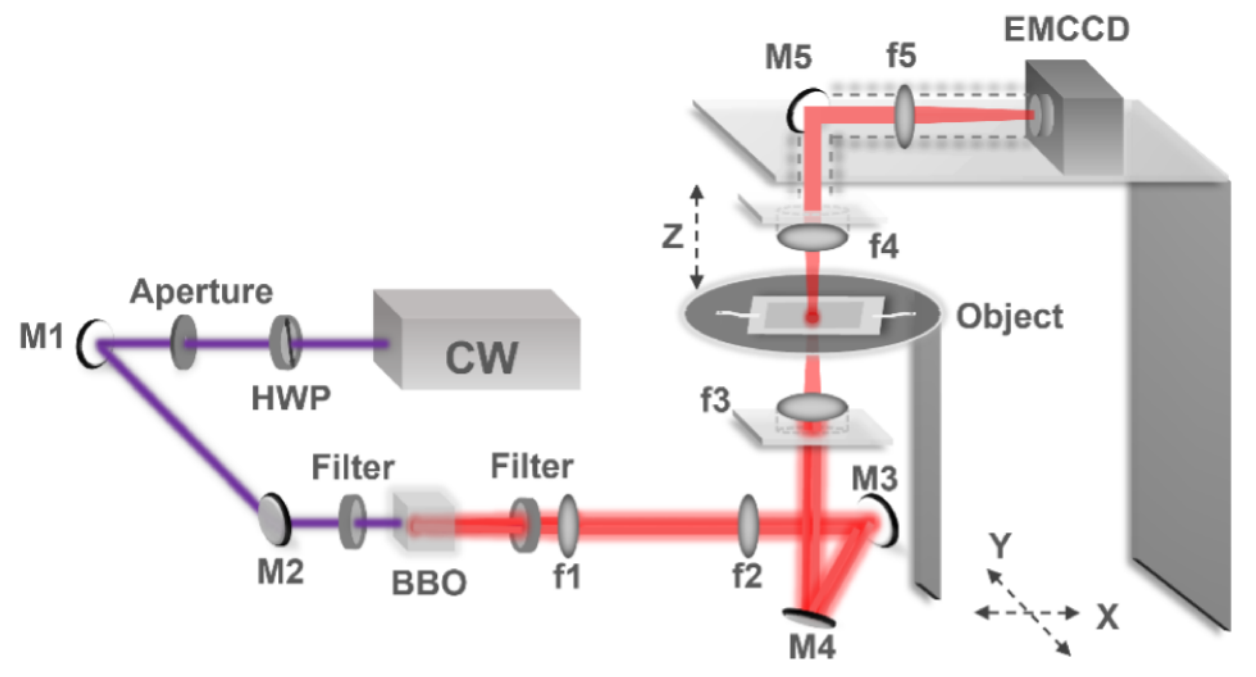

**Figure 1.** Experimental setup for camera-based quantum imaging with SPDC. A 405 nm CW laser beam is reflected by mirrors M1 and M2 through an aperture and half-wave plate (HWP) before entering a $\beta$-BBO crystal for Type-I spontaneous parametric down-conversion, generating degenerate photon pairs at 810 nm. The collinear signal and idler beams pass through spectral filters and lenses f1 (50 mm) and f2 (100 mm) for wavelength selection, then are reflected by mirrors M3 and M4 upward into the imaging path. The beam transmits through a USAF resolution target (Object) mounted on an X-Y translation stage, positioned between lenses f3 and f4 (both 50 mm focal length). The two lenses can be replaced by objectives. The transmitted light is reflected by M5 and passes through lens f5 (100 mm) before reaching an EMCCD camera. This 4f imaging configuration enables spatially resolved detection of correlated photon pairs for quantum imaging.

To identify candidate pixel pairs, we applied geometric preselection using approximate central symmetry within a 70-pixel search radius centered on the intensity maximum. This radius exceeds the SPDC correlation region, which extends to approximately 50 pixels as evidenced by the saturation of correlation coefficient and covariance in Figure S3. For statistical discrimination analysis, we applied spatial masking to exclude pixels blocked by the USAF target and their symmetric pairs. This masking restricts the analysis to the SPDC-illuminated region, demonstrating that $\delta$ discriminates correlated pixel pairs rather than artifacts from blocked or unilluminated areas. For image reconstruction, all pixels within the search region were evaluated.

For each pixel $i$, we computed the excess kurtosis from its temporal intensity series and calculated the covariance and kurtosis difference $\delta$ for each candidate pair $(i, j)$. The kurtosis-weighted reconstruction computes the intensity at each position as $I_{\mathrm{recon},i}$ using eq 9. All statistical quantities are evaluated from the raw, baseline-subtracted intensity before any image normalization or interpolation. Min−max normalization is applied only to the final reconstructed images for display. The line profiles in Figure 4a are extracted from the normalized images and displayed with cubic-spline interpolation.

KWC has a higher computational cost than the single-center covariance method per frame. COV evaluates the covariance only for the $\sim N_{\mathrm{pixels}}$ mirror pairs about a preselected correlation center, at $O(N_{\mathrm{pixels}} \times N_{\mathrm{frames}})$, while KWC evaluates all $N_{\mathrm{cand}}$ candidate partners per pixel within the search region, at $O(N_{\mathrm{pixels}} \times N_{\mathrm{cand}} \times N_{\mathrm{frames}})$. As acquisition, frame loading, kurtosis evaluation, and covariance accumulation all scale linearly with the number of frames, the 40-fold reduction in the number of frames lowers the end-to-end cost despite the higher per-frame cost.

## RESULTS

We first examined whether kurtosis difference discriminates correlated pixel pairs from random combinations. Candidate pairs, preselected by central symmetry, show mean $\langle\delta\rangle_{\mathrm{pixel}}$ = 9.34, while random controls have $\langle\delta\rangle_{\mathrm{pixel}}$ = 44.39, making a 4.75-fold separation, as illustrated in Figure 2a. The symmetric pixel distribution is sharply peaked near zero with a fast-decreasing tail, while the asymmetric pixel distribution is broad and relatively flat. This separation arises because symmetric pixels share similar rare-event statistics. The histogram of absolute kurtosis difference for all pixels is presented in Figure S4. Figure 2b shows kurtosis values for symmetric and asymmetric pixels. Symmetric pixels cluster along the diagonal $\kappa_i = \kappa_j$ with strong correlation (Pearson $r \approx 0.85$), while asymmetric pixels scatter broadly with weak correlation ($r \approx 0.15$). Repeating results using a different object are shown in Figure S5.

To quantify the discrimination performance, we performed receiver operating characteristic (ROC) analysis comparing kurtosis difference and Pearson correlation coefficient as classifiers for identifying symmetric pixels, as shown in Figure 2c. The ROC curve shows the true positive rate (correctly identified symmetric pixels) against the false positive rate (asymmetric pixels misclassified as symmetric) across all possible classification thresholds. Absolute kurtosis difference achieves an area-under-curve (AUC) of 0.830, substantially outperforming Pearson correlation with an AUC of 0.658. An AUC of 1.0 represents perfect classification, while 0.5, shown by the dashed line, indicates random guessing. The 26% improvement in AUC confirms that our approach provides

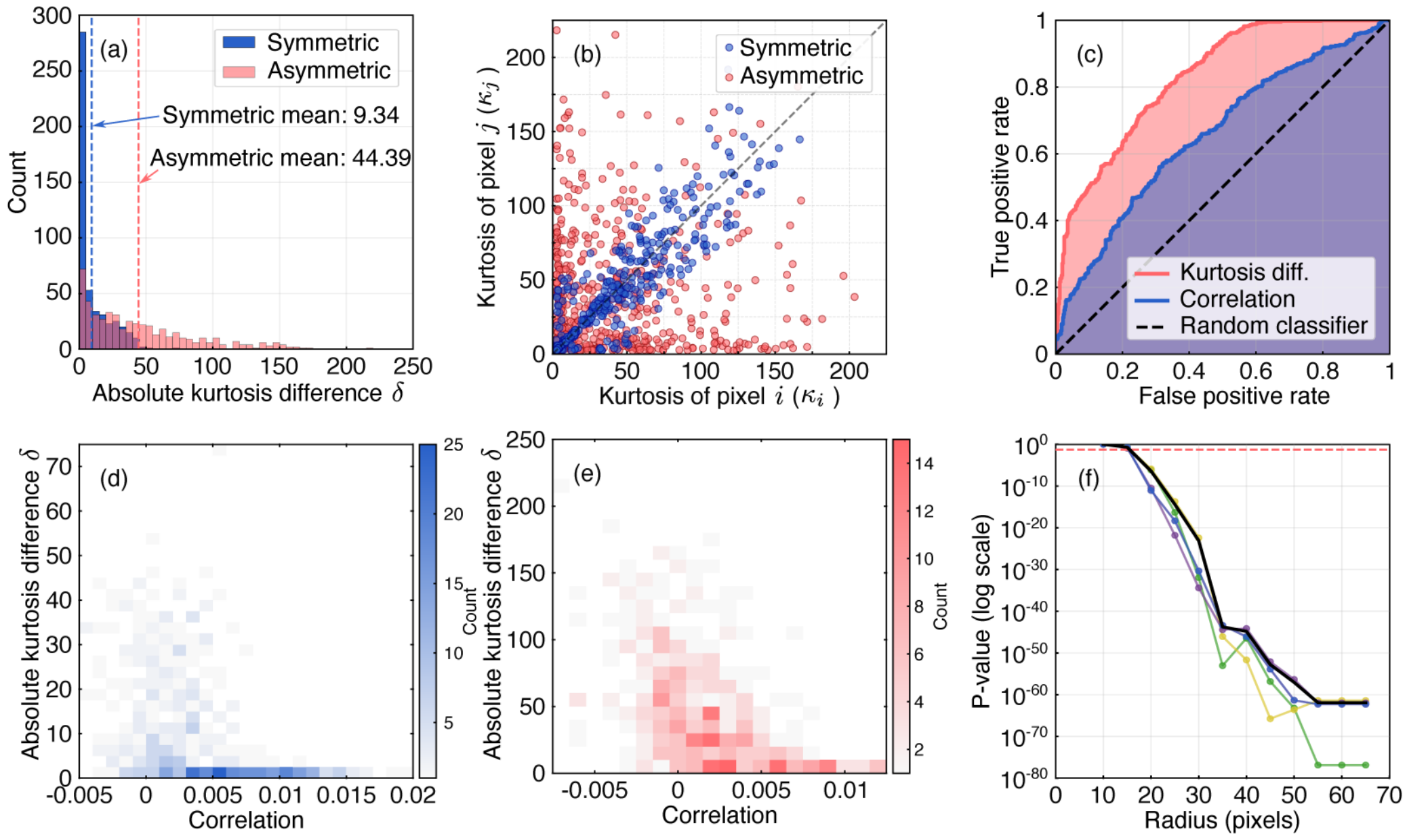

**Figure 2.** Statistical discrimination using kurtosis difference. (a) Histogram of $\delta$ for 500 symmetric pixels (blue, selected by central symmetry) and 500 asymmetric pixels (red, randomly selected). Symmetric pixels concentrate at small $\delta$ ($\langle\delta\rangle_{\text{pixel}} = 9.34$), while asymmetric pixels show broad distribution ($\langle\delta\rangle_{\text{pixel}} = 44.39$), demonstrating a 4.75-fold separation. (b) Scatter plot of kurtosis values ($\kappa_i$ vs $\kappa_j$) for symmetric and asymmetric pixel pairs. Symmetric pixels cluster along the diagonal $\kappa_i = \kappa_j$ (indicating similar tail statistics), while asymmetric pixels scatter broadly. (c) Receiver operating characteristic curves comparing discrimination performance of kurtosis difference [red, area-under-curve (AUC) = 0.830] versus Pearson correlation coefficient (blue, AUC = 0.658) for identifying symmetric pixels. Kurtosis difference provides superior discrimination, with both metrics significantly outperforming random classification (dashed line, AUC = 0.5). (d, e) Joint distribution of correlation coefficient and $\delta$ for 500 symmetric (d) and asymmetric (e) pixels. Asymmetric pixels exhibit broad $\delta$ distribution with near-zero correlation, showing no clear relationship between the two metrics. Symmetric pixels cluster at low $\delta$ despite weak correlation coefficients, confirming that tail-statistics matching provides discrimination where intensity correlation fails. (f) Statistical significance test: p-value for Mann−Whitney U test comparing $\delta$ distributions of symmetric versus asymmetric pixels as a function of search radius (10−65 pixels) around the SPDC center. Five colored curves show independent trials; the black curve shows their mean. For radii exceeding 15 pixels, $p < 0.05$, confirming robust discrimination across the SPDC illumination region.

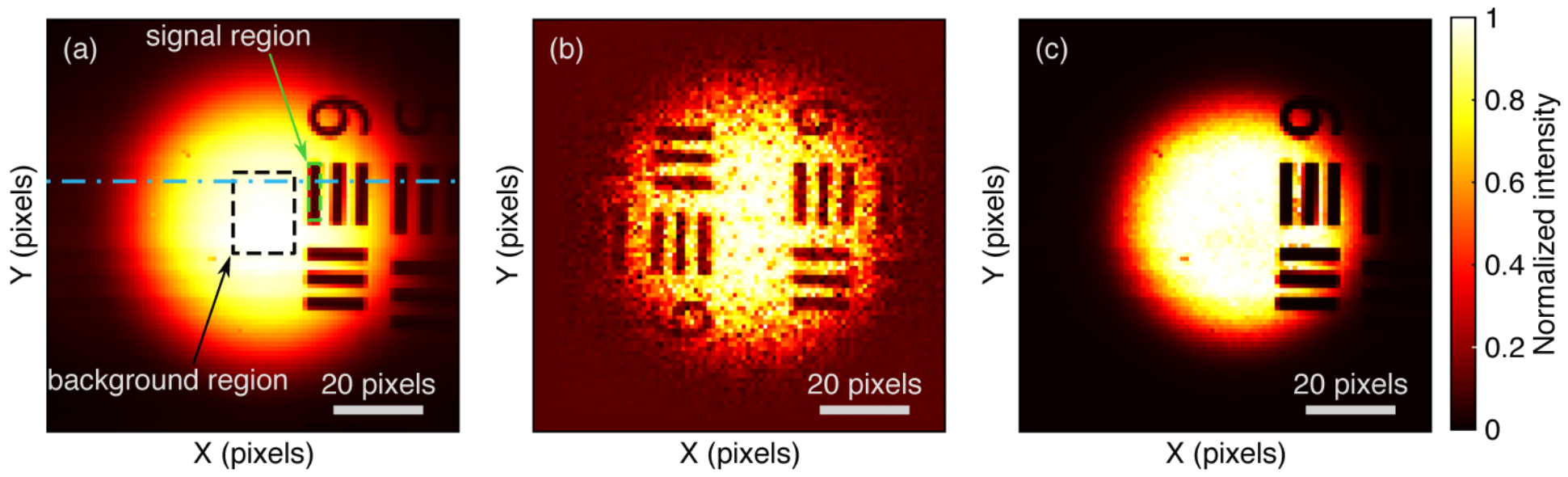

**Figure 3.** Reconstruction comparison for a USAF resolution target using three methods. (a) Accumulation (ACC). Direct summation of all detected photons. Dashed rectangles indicate background region (black dashed box) and signal region (green dashed box) used for contrast-to-noise ratio (CNR) calculation. The cyan dash-dot line marks the position of the line profiles analyzed in Figure 4a. (b) Standard covariance (COV). Reconstruction using central-symmetric pixel pairs without weighting, showing elevated background noise. (c) Kurtosis-weighted covariance (KWC). Reconstruction using $\exp(-\delta/\kappa_0)$ weighting to suppress noise from pairs with mismatched tail statistics, achieving superior background suppression and edge definition.

more reliable pair identification than intensity correlation for this data set.

Figure 2d−e shows how kurtosis difference and correlation coefficient distribute across pixel pairs. 5000 frames are used for this examination. Asymmetric pixels have larger variance in $\delta$ because their kurtosis values fluctuate independently. Photons at these positions come from different emission events with unrelated statistics. Symmetric pixels have much smaller variance of $\delta$ because they detect photons from the same emission position. The variance behaves differently at low correlation coefficients. Here, intensity correlation over frames cannot distinguish symmetric from asymmetric pixels, but kurtosis difference can. This pattern motivates the weighted reconstruction approach, where small $\delta$ indicates high pairing probability.

We tested kurtosis discrimination across different spatial scales to validate this approach. For circular regions with radii from 10 to 65 pixels around the SPDC center, we selected 500 centrally symmetric pixels and 500 randomly asymmetric pixels, then applied the Mann−Whitney U test to check whether symmetric pixels have smaller $\delta$. Five independent trials (Figure 2f) show the same pattern. P-values drop below 0.05 for radii above 15 pixels, which means discrimination is statistically significant across the SPDC illumination region. Below 15 pixels, small sample size and limited spatial diversity reduce statistical power. Beyond 50 pixels, p-values plateau, consistent with saturation in correlation metrics (Figure S3). The SPDC correlation region extends to about 50 pixels.

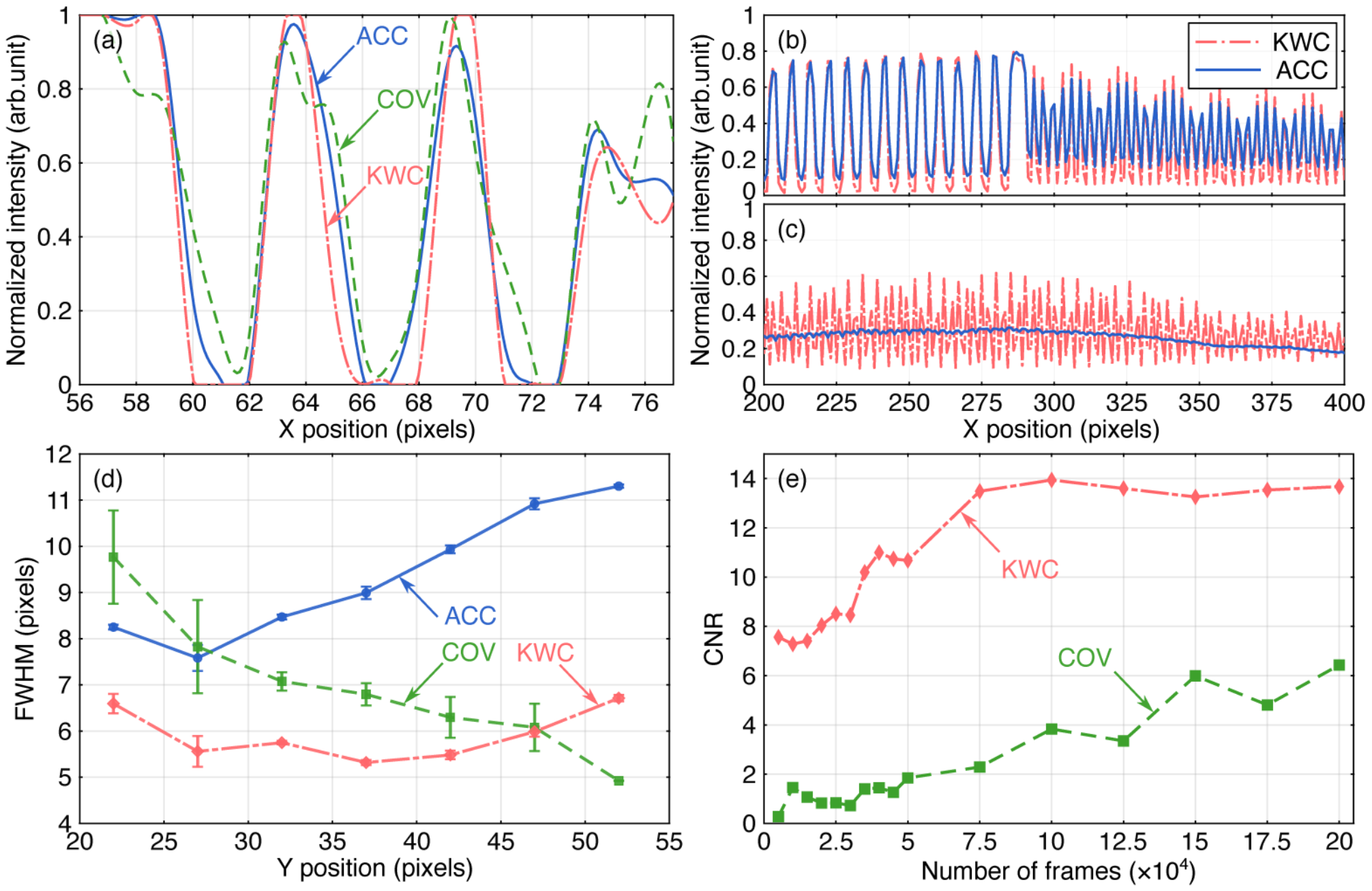

**Figure 4.** Quantitative comparison of reconstruction methods. (a) Normalized intensity line profiles through USAF target bars indicated by the thick cyan dash-dot line in Figure 3a. All profiles are min−max normalized for direct comparison and displayed with cubic-spline interpolation. KWC (red, dot-dashed) shows sharper edge transitions and lower baseline noise compared to COV (green, dashed) and ACC (blue, solid). (b, c) Cross-line profiles through the bar field of the periodic bar targets, over the bar rows of the normalized reconstructions shown in Figure S9, ACC (blue) versus KWC (red). (b) The target with the low- and medium-frequency groups. Both methods resolve the bar comb, and KWC reaches deeper minima, leading to improved CNR. (c) The high-frequency target. The ACC profile is essentially flat (contrast 0.06) while the KWC profile retains the full bar comb (contrast 0.74). (d) Spatial resolution measured by fwhm across 14 vertical positions shown in Figure S7. Kurtosis-weighted covariance achieves 5.92 ± 0.51 pixels, a 37% improvement over accumulation and 15% over standard covariance. (e) Frame efficiency analysis shows kurtosis-weighted covariance reaches CNR > 7 at 5000 frames, matching the performance that standard covariance achieves at 200000 frames, a 40-fold reduction in acquisition time.

Figure 3 compares three reconstruction methods applied to a USAF resolution target. Direct accumulation (ACC, Figure 3a) sums all detected photons without using SPDC correlations. Standard covariance (COV, Figure 3b) uses mirror symmetric pixel pairs about a preselected center. Covariance physically distinguishes true coincidences from accidental ones by counting how often photon pairs arrive together. The kurtosis weight identifies which pairs are likely correlated but says nothing about coincidence strength. Kurtosis-weighted covariance (KWC, Figure 3c) uses both. Pairs with similar sparse-photon statistics and strong coincidence counts get more weight, and weakly correlated pairs get suppressed. The weighting is soft, not a hard cutoff. This boosts contrast-to-noise ratio (CNR, defined in Supplementary Note 3) by a factor of 2.29 compared to unweighted covariance and produces cleaner images with less noise and sharper features. The comparison of COV and KWC with different frames is shown in Figure S6.

Figure 4a compares normalized intensity line profiles across the USAF target bars for images in Figure 3. ACC produces gradual edge transitions typical of diffraction-limited imaging. COV enhances spatial resolution using photon pair correlations but introduces baseline noise. KWC improves the SNR of COV while enhancing the spatial resolution. The line profiles with different frames are shown in Figure S8.

We measured full width at half-maximum (FWHM) from edge spread functions at vertical positions in Figure S7. As shown in Figure 4d, ACC gives fwhm = 9.35 ± 1.30 pixels, which stands for the classical limit. COV improves this to 6.97 ± 1.42 pixels (25% better). KWC reaches 5.92 ± 0.51 pixels, 37% better than ACC and 15% better than COV. While quantum imaging theory predicts a 2-fold resolution enhancement[4,6,9], the observed improvement falls short of this limit for two reasons. First, correlated photon pair events in SPDC occur far less frequently than photon detection in classical imaging, resulting in weaker edge signals and reduced spatial resolvability compared to the theoretical 2-fold limit. Second, finite-frame statistics cause a fraction of asymmetric pixels to have small $\delta$ by chance, producing residual weights that slightly broaden reconstructed edges. Both effects diminish with increasing frame count, suggesting that longer acquisition approaches the theoretical limit.

We repeat resolution measurements on the periodic-bar target with a contrast analysis that does not rely on edge fitting. For a periodic target the contrast is $C = (A_{max} - A_{min})/(A_{max} + A_{min})$, with $A_{max}$ and $A_{min}$ the maximum and minimum values of the bar profile, measured only in the central region where the illumination is strongest. KWC shows better contrast than ACC at every bar frequency, with the advantage growing toward higher frequencies. The contrasts of ACC and KWC fall from 0.75 and 0.93 for the low-frequency group to 0.60 and 0.84 for the medium-frequency group. On the separate high-frequency results, the contrast is 0.06 for ACC against 0.74 for KWC. Therefore, the KWC reconstruction of the finer target retains the full bar comb where the ACC image is essentially flat (Figure 4c and Figure S9).

CNR analysis shows clearer noise suppression. At 200000 frames, ACC yields CNR = 12.87 with signal 0.940 ± 0.003 and background 0.076 ± 0.002. COV achieves only CNR = 6.43 as background rises to 0.155, indicating substantial accidental correlations. KWC reaches CNR = 13.68, 2.1 × better than COV. Signal (0.952 ± 0.003) matches ACC while background drops to 0.0320 ± 0.0001. Frame efficiency analysis (Figure 4e) shows KWC reaches CNR > 7 at 5000 frames and rises to 13.68 at 200000 frames. COV stays below CNR = 2.5 with less than 75000 frames and needs 200000 frames to match what KWC achieves at 5000 frames, a 40-fold reduction in acquisition time.

The higher CNR of KWC relative to ACC originates in a higher image contrast. Usually, the classical imaging is not shot-noise-limited, because the ACC also contains uncorrelated noise from stray light and detector noise. Correlation-based detection rejects these uncorrelated contributions, which lowers the noise background and raises the region contrast of the covariance-based image above that of ACC. This behavior is well established in quantum imaging.[5,8,9,16,17]

## DISCUSSION

Kurtosis difference enables quantum imaging without calibrating correlation centers. The approach exploits how correlated pixels share similar rare event statistics. These pixels produce heavy-tailed intensity distributions with comparable kurtosis. Uncorrelated pixels detect photons from independent emission positions, giving distinct tail statistics and larger kurtosis differences. Weighting covariance by $\exp(-\delta/\kappa_0)$ emphasizes pairs with similar photon statistics and suppresses dissimilar pairs.

KWC outperforms COV because of how they select pixel pairs. COV uses a single preselected correlation center and measures covariance only for pairs symmetric about that center. This works when the crystal has one dominant correlation center, but thick crystals produce multiple centers (Figure S2).[10,11] COV can only exploit correlations related to the preselected center and ignores correlations related to other centers. KWC automatically identifies correlated pairs across the entire search region using kurtosis difference, then calculates their covariance. This captures correlations from all emission positions without requiring center selection. At 200000 frames, COV shows less distinguishable signal (0.155 vs 0.076 for ACC) because it includes uncorrelated pairs forced by single-center geometry while missing true pairs symmetric about other centers. KWC reaches CNR = 13.68 with background at 0.032 by finding correlated pairs wherever they occur. COV performs worse than ACC because single-center geometry is too restrictive. Kurtosis weighting solves this problem by selecting pairs by tail statistics across all candidate pixels, then measuring their covariance.

The method has limits as well. Random fluctuations sometimes produce small $\delta$ even for asymmetric pixels (overlapping heads in Figure 2a). Weighting is soft, not a hard cutoff. Small $\delta$ means higher weight whether the pair is truly correlated or not. The covariance term ensures only pairs with real coincidence counts contribute much. We investigate all candidate pixel pairs in a broad search region, which handles multiple centers and optical distortions without preselected centers. We have not benchmarked against classical correlated light, and detector artifacts could affect kurtosis systematically. Reliable kurtosis needs enough frames; performance improves from 5000 to 200000 frames (Figure 4e). Parameter optimization across flux regimes, extension to Type-II SPDC and nondegenerate wavelengths, and dynamic imaging remain open questions.

## CONCLUSION

Absolute kurtosis difference identifies spatially correlated photon pairs in camera-based quantum imaging without preselecting correlation centers. Correlated pixels detecting photons from the same SPDC emission position share similar rare-event statistics. They produce heavy-tailed distributions with matched kurtosis. Uncorrelated pixels from independent SPDC emission positions show distinct tail statistics. Standard covariance uses a single preselected center and misses correlations from other emission positions in thick crystals. Kurtosis weighting finds correlations wherever they occur, handling multiple emission positions. As a result, weighting covariance by $\exp(-\delta/\kappa_0)$ exploits this difference to improve CNR and reduce acquisition time by 40 times.

## ASSOCIATED CONTENT

### Data Availability Statement

The raw camera data and the analysis code that support this study are available from the corresponding author upon reasonable request.

### Supporting Information

The Supporting Information is available free of charge at https://pubs.acs.org/doi/10.1021/acsphotonics.6c01982.

Derivation of the variance of the kurtosis difference for symmetric and asymmetric pixel pairs; derivation of the expected absolute kurtosis difference; definition of the contrast-to-noise ratio; Monte Carlo simulation of kurtosis-difference discrimination; correlation-center maps for SPDC; dependence of correlation metrics on radius; histogram of absolute kurtosis difference over all pixel pairs; results on additional data; reconstructed images versus frame count; USAF target alignment with fwhm measurement positions; line profiles with different frame counts; periodic bar target reconstructions (Figures S1−S9) (PDF)

## AUTHOR INFORMATION

### Corresponding Author

**Zhe He** − *Shandong Institute of Advanced Technology, Jinan, Shandong 251000, China*; orcid.org/0000-0002-8525-3650; Email: zhe.he@iat.cn

### Authors

**Yanli Shi** − *Shandong Institute of Advanced Technology, Jinan, Shandong 251000, China*

**Hui Wu** − *Shandong Institute of Advanced Technology, Jinan, Shandong 251000, China*

**Qun Cao** − *Shandong Institute of Advanced Technology, Jinan, Shandong 251000, China*; orcid.org/0000-0002-8727-122X

**Weidong Zheng** − *Shandong Institute of Advanced Technology, Jinan, Shandong 251000, China*; orcid.org/0000-0002-9793-5763

**Zheng Cui** − *Shandong Institute of Advanced Technology, Jinan, Shandong 251000, China*

Complete contact information is available at:
https://pubs.acs.org/10.1021/acsphotonics.6c01982

## Author Contributions

#Z.H. and Y.S. contributed equally to this work. Z.H. conceived the project and experiment. Y.S. contributed to the imaging reconstruction theory. Z.H., Y.S., H.W., Q.C., W.Z., and Z.C. participated in the discussion, analysis, and manuscript preparation.

## Funding

The project is supported by the National Natural Science Foundation of China under grant No. 12404413, the Natural Science Foundation of Shandong Province (Nos. 2024HWYQ-081 and ZR2026MS0050), and the Taishan Scholars Foundation of Shandong Province (Nos. tsqn202408307 and tsqn202306316).

## Notes

A preprint version of this work is available: Zhe He; Yanli Shi; Hui Wu; Qun Cao; Weidong Zheng; Zheng Cui. Quantum Imaging via Kurtosis-Difference Weighted Covariance on a 2D Camera. 2026, arXiv:2606.31005. https://arxiv.org/abs/2606.31005 (accessed October 14, 2026).

The authors declare no competing financial interest.

## ■ REFERENCES

(1) Moreau, P.-A.; Toninelli, E.; Gregory, T.; Padgett, M. J. Imaging with Quantum States of Light. *Nat. Rev. Phys.* **2019**, *1*, 367−380.

(2) Pearce, E.; Nothlawala, F.; Forbes, A.; Padgett, M. J. Quantum Imaging with Correlated Photon Pairs. *Nat. Rev. Methods Primers* **2026**, *6*, 17.

(3) Yue, X.; Wu, H.; Wang, J.; He, Z. Quantum Super-Resolution Imaging: A Review and Perspective. *Nanophotonics* **2025**, *14*, 1961−1974.

(4) Defienne, H.; Cameron, P.; Ndagano, B.; Lyons, A.; Reichert, M.; Zhao, J.; Harvey, A. R.; Charbon, E.; Fleischer, J. W.; Faccio, D. Pixel Super-Resolution with Spatially Entangled Photons. *Nat. Commun.* **2022**, *13*, 3566.

(5) Brida, G.; Chekhova, M. V.; Fornaro, G. A.; Genovese, M.; Lopaeva, E. D.; Berchera, I. R. Systematic Analysis of Signal-to-Noise Ratio in Bipartite Ghost Imaging with Classical and Quantum Light. *Phys. Rev. A* **2011**, *83*, 063807.

(6) Defienne, H.; Ndagano, B.; Lyons, A.; Faccio, D. Polarization Entanglement-Enabled Quantum Holography. *Nat. Phys.* **2021**, *17*, 591−597.

(7) Edgar, M. P.; Tasca, D. S.; Izdebski, F.; Warburton, R. E.; Leach, J.; Agnew, M.; Buller, G. S.; Boyd, R. W.; Padgett, M. J. Imaging High-Dimensional Spatial Entanglement with a Camera. *Nat. Commun.* **2012**, *3*, 984.

(8) Gregory, T.; Moreau, P.-A.; Toninelli, E.; Padgett, M. J. Imaging through Noise with Quantum Illumination. *Sci. Adv.* **2020**, *6*, No. eaay2652.

(9) He, Z.; Zhang, Y.; Tong, X.; Li, L.; Wang, L. V. Quantum Microscopy of Cells at the Heisenberg Limit. *Nat. Commun.* **2023**, *14*, 2441.

(10) Septriani, B.; Grieve, J. A.; Durak, K.; Ling, A. Thick-Crystal Regime in Photon Pair Sources. *Optica* **2016**, *3*, 347−350.

(11) He, Z.; Wu, H.; Li, G.; Shao, C.; Zheng, W.; Cao, Q.; Cui, Z. Quantum Super-resolution Imaging with Multicenter Integration. *ACS Photonics* **2025**, *12*, 6530−6534.

(12) Wolley, O.; Gregory, T.; Beer, S.; Higuchi, T.; Padgett, M. Quantum Imaging with a Photon Counting Camera. *Sci. Rep.* **2022**, *12*, 8286.

(13) Blauensteiner, B.; Herbauts, I.; Bettelli, S.; Poppe, A.; Hübel, H. Photon Bunching in Parametric Down-Conversion with Continuous-Wave Excitation. *Phys. Rev. A* **2009**, *79*, 063846.

(14) Schneeloch, J.; Howell, J. C. Introduction to the Transverse Spatial Correlations in Spontaneous Parametric Down-Conversion through the Biphoton Birth Zone. *J. Opt.* **2016**, *18*, 053501.

(15) Büse, A.; Tischler, N.; Juan, M. L.; Molina-Terriza, G. Where Are Photons Created in Parametric Down-Conversion? On the Control of the Spatio−Temporal Properties of Biphoton States. *J. Opt.* **2015**, *17*, 065201.

(16) Tasca, D. S.; Edgar, M. P.; Izdebski, F.; Buller, G. S.; Padgett, M. J. Optimizing the Use of Detector Arrays for Measuring Intensity Correlations of Photon Pairs. *Phys. Rev. A* **2013**, *88*, 013816.

(17) Lopaeva, E. D.; Ruo Berchera, I.; Degiovanni, I. P.; Olivares, S.; Brida, G.; Genovese, M. Experimental Realization of Quantum Illumination. *Phys. Rev. Lett.* **2013**, *110*, 153603.